\documentstyle[prl,aps,psfig]{revtex}

\begin{document}
\draft

\twocolumn[\hsize\textwidth\columnwidth\hsize\csname
@twocolumnfalse\endcsname

\widetext
\title{Antiferromagnetic integer-spin chains in a staggered magnetic field: 
approaching the thermodynamic limit through the infinite-size DMRG.}
\author{Massimo Capone and Sergio Caprara}  
\address{ Dipartimento di Fisica, Universit\`a di Roma ``La Sapienza'',
and  Istituto Nazionale per la Fisica della Materia (INFM),
Unit\`a  di Roma 1, Piazzale Aldo Moro 2, I-00185 Roma, Italy}
\date{\today}
\maketitle

\begin{abstract}
We investigate the behavior of antiferromagnetic integer-spin chains
in a staggered magnetic field, by means of the density-matrix
renormalization group, carefully addressing 
the role of finite-size effects within the Haldane phase at small fields.
In the case of spin $S=2$, we determine the dependence of the groundstate 
energy and magnetization on the external field, in the thermodynamic limit,
and show how the peculiar finite-size behavior can be connected with the
crossover in the groundstate from a spin liquid to a polarized N\'eel state.
\end{abstract}

\pacs{75.10.Jm, 75.30.Cr, 75.40.Mg}
]

\narrowtext

\section{Introduction}
\label{intro}
The behavior of antiferromagnetic (AFM) Heisenberg spin chains is dominated by
quantum fluctuations which suppress magnetic long-range order. However, after 
Haldane's proposal \cite{hal}, it has become clear that the value of the spin 
$S$ plays a crucial role. Integer and half-integer spin cases are indeed 
completely different. The spectrum of the half-integer spin chain is gapless 
in the thermodynamic limit, and the system is critical, in the sense that the 
linear response to an infinitesimal staggered magnetic field, coupled to the 
would-be order parameter, is divergent (at zero temperature). Conversely, the 
integer-spin chain has a spectrum which stays gapped in the thermodynamic 
limit. The first excited state has a finite distance from the groundstate 
(the so-called Haldane gap), and the system is not critical, i.e., the linear 
response to an infinitesimal staggered magnetic field is finite. The 
groundstate, which is characterized by a finite correlation length, is a spin 
liquid.

The two behaviors are only reconciled as $S\to \infty$. This limiting case,
which corresponds to the suppression of quantum fluctuations, leads to the 
closure of the Haldane gap and to the divergence of the staggered 
susceptibility even in the integer-spin case, which becomes thus 
indistinguishable from the half-integer-spin case, as it is naively expected
at large $S$. 

All theoretical approaches are well controlled only in the limit of large $S$,
while the increasing relevance of quantum fluctuations makes the predictions
less accurate as $S$ is reduced. On the other hand, the introduction of the 
staggered magnetic field gradually freezes quantum fluctuations, and the 
problem arises of the description of the spin-liquid state in a staggered 
magnetic field and its evolution to the frozen state. A non-linear 
$\sigma$-model approach has been recently developed to investigate this 
evolution \cite{bolognesi}. This approach relies on the Haldane {\it ansatz}, 
and becomes exact in the limit $S\to\infty$, in the absence of staggered 
magnetic field. However, the validity of the {\it ansatz} must be limited by 
some finite field, at which the spin-liquid description breaks down. Moreover, 
at small $S$, the starting {\it ansatz} is (quantitatively) not appropriate 
even at zero field, and corrections must be considered, when describing the 
field-driven crossover from the spin liquid to the polarized AFM state. 

This paper is therefore devoted to the numerical analysis of the response of 
the AFM integer-spin chains to a staggered magnetic field, by means of the
density-matrix renormalization group (DMRG) \cite{white}, which 
gives basically exact results for one-dimensional systems, that are not 
biased by any {\it a priori} assumption, and represent an almost ideal 
numerical test for analytical results.

The case $S=1$ has been most extensively studied within DMRG\cite{s1}, 
but the case $S=2$ \cite{s2} represents quite a hard numerical task even 
in the absence of a magnetic field, and lacks any analysis in the presence 
of the staggered field, that will be the main object of our numerical 
investigation.

Our starting Hamiltonian reads
\begin{equation}
\label{hamiltonian}
{\cal H} = J\sum_{i=1}^{L-1} {{\bf S}_i} \cdot {{\bf S}_{i+1}} -
H \sum_{i=1}^L (-1)^{i}S_i^z,
\end{equation}
where ${\bf S}_i$ is the spin-$S$  operator on site $i$, $S=1,2,...$ is 
integer, and $L$ is the number of sites in the chain. $J > 0$ is the AFM 
Heisenberg coupling, $H$ is the amplitude of the staggered magnetic field 
along the $z$ axis, which is coupled with the staggered magnetization 
$M^z = \sum_i (-1)^i S_i^z$. In view of the forthcoming DMRG study, we 
explicitly assumed open boundary conditions (OBC) in writing the first term 
in (\ref{hamiltonian}). 

The Hamiltonian (\ref{hamiltonian}) is manifestly not invariant under
reflection with respect to the mid point of the chain when $L$ is even.
However, in the standard implementation of the DMRG algorithm \cite{white},  
$L$ is even, and the reflection symmetry is used to reduce the numerical 
effort by identifying the right block with the reflection of the left one 
(for more details, see the discussion in Sec. \ref{convergency}). Although 
there is no problem in implementing a non-symmetric algorithm \cite{ising,xy}, 
the Hamiltonian (\ref{hamiltonian}) can be easily made invariant under 
reflection by means of a simple gauge transformation (i.e. a local rotation 
of the reference frame), given by $(-1)^i S_i^z \rightarrow {\tilde S}_i^z$, 
and, e.g., $(-1)^i S_i^x \rightarrow {\tilde S}_i^x$, while ${\tilde S}_i^y$
coincides with  $S_i^y$. Once the transformation is performed, the 
Hamiltonian (\ref{hamiltonian}) is re-cast in the form
\begin{equation}
\label{gauged}
{\tilde {\cal H}} = - J\sum_{i=1}^{L-1}
\left[ {\tilde S}_i^z {\tilde S}_{i+1}^z +
{1\over 2}\left( {\tilde S}_i^{+}{\tilde S}_{i+1}^{+}
 + {\tilde S}_i^{-}{\tilde S}_{i+1}^{-} \right)\right] -
 H \sum_{i=1}^L {\tilde S}_i^z, 
\end{equation}
where ${\tilde S}_i^{\pm}={\tilde S}_i^{x}\pm {\rm i}{\tilde S}_i^{y}$,
and the external field is now coupled to the uniform magnetization
${\tilde S}^z = \sum_i {\tilde S}_i^z$. The token we pay to recover 
reflection symmetry is the appearance of terms of the form 
$S_i^{+}S_{i+1}^{+}$ and $S_i^{-}S_{i+1}^{-}$, in which the spin is
simultaneously raised or lowered on neighboring sites. 
 
The original Hamiltonian (\ref{hamiltonian}) commutes with the $z$-component
of the total spin $S^z = \sum_i S_i^z$ and, therefore, the gauge-transformed 
Hamiltonian (\ref{gauged}) commutes with the staggered magnetization
${\tilde M}^z = \sum_i (-1)^i {\tilde S}_i^z$. It is easily realized that the 
groundstate of the model lies within the ${\tilde M}^z = 0$ subspace at large 
values of the field $H$. At zero field, the groundstate is instead degenerate 
(in the presence of OBC), but, since the total spin is obviously integer, a 
state with ${\tilde M}^z = 0$ is certainly  present in the groundstate 
multiplet. We explicitly checked that (at least a component of) the
groundstate always lies in the ${\tilde M}^z=0$ subsector for every value of 
the amplitude of the external field $H$. Henceforth, for the sake of 
definiteness, we assume $H>0$, the case of negative $H$ being trivially 
recovered by changing ${\tilde S}^z\rightarrow -{\tilde S}^z$ and 
${\tilde M}^z\rightarrow -{\tilde M}^z$ in the following discussion. In 
particular the groundstate energy and magnetization are even and odd 
functions of the external field respectively. In the rest of the paper we 
refer to the properties of the model described by (\ref{gauged}), and in 
particular to the uniform magnetization along the external field, 
$\langle {\tilde S}^z\rangle$. The correspondence with the original model 
(\ref{hamiltonian}) is straightforward.

The plan of the paper is the following: in Sec. \ref{convergency} we discuss 
the general problem of extracting the thermodynamic properties within the 
DMRG approach, in connection with the interplay of finite-size effects and 
truncation of the Hilbert space, and we specialize the discussion to the case 
of the Haldane spin-liquid phase at small magnetic field; in Sec. 
\ref{results} we present our numerical results, mostly on the $S=2$ case, and 
discuss the various regimes which characterize the model at small (Sec. 
\ref{piccoli}) and large (Sec. \ref{largefields}) magnetic field, showing how 
these can be also characterized by the peculiar form of the finite-size 
corrections to the magnetization; concluding remarks are found in Sec. 
\ref{concl}.

We point out that throughout the paper, whenever we discuss finite-size 
corrections, we specifically refer to corrections in the presence of OBC, 
which are those affecting our DMRG results.

\section{Extracting the Thermodynamic properties}
\label{convergency}

After the introduction, by S. R. White \cite{white}, of the DMRG, a wide 
majority of the DMRG studies focused on the so-called finite-size algorithm, 
in which the system size is first increased until a chosen size is reached, 
and then stopped. Subsequently, a few further ``sweeps'' are performed to 
improve the basis for that system size. On the other hand, in the 
infinite-size algorithm the size is always increased at each step. It is 
clear that, for a given lattice size, the finite-size algorithm gives the 
best estimate (at fixed Hilbert space). Conversely, if one is really 
interested in the thermodynamic limit, the information obtained with the 
infinite-size algorithm are of extreme value \cite{caprara,marston}, and 
become crucial in the integer-spin case at issue in this paper, as we show
below. We point out that extreme care has to be taken in checking the
convergence of the procedure with respect to the truncation of the Hilbert 
space, before any sensible statement on the physical meaning of the numerical 
results is made.

The algorithm we used is the standard implementation of the infinite-size 
DMRG, which is described in \cite{white} (see also \cite{ising,xy}). Here we 
limit ourselves to a quick summary which serves as an introduction to the 
notations which we use in the paper. As it is customary we start with a 
4-site chain which is divided into a left and a right part (blocks). The 
density matrix of, say, the right block in the groundstate of the entire 
system (superblock) is diagonalized and $N_k$ eigenstates, corresponding to 
the largest eigenvalues, are kept as the most representative of the block, 
while the other eigenstates are truncated away, thus reducing the size of the
Hilbert space. In the presence of reflection 
symmetry with respect to the mid point of the chain, the left block is simply 
the reflection of the right block. Once the truncated left and right blocks 
are obtained, two more sites are added in the middle of the chain, and the 
new left and right blocks are defined, each as the ``old truncated block + 
newly added site'' system. The procedure is then iterated. After $N_{RG}$ 
iterations the chain contains $L=2N_{RG}+2$ sites. At each step, a measure of 
the reliability of the procedure is provided by the truncation error 
${\cal R}=1-\sum_{i=1}^{N_k} w_i$, where $w_i$ are the eigenvalues of the 
density matrix, sorted in decreasing order. It follows from the definition 
that $0 \le {\cal R}\le 1$, and that, in the most general case, ${\cal R} = 1$
only when all the basis states are kept at each iteration, unless, under
special circumstances, a simplification occurs in the spectrum of the density
matrix (see below, Sec. \ref{largefields}). The success of DMRG when dealing
with one-dimensional systems is due to their peculiar topology in the presence
of short-range interactions. In most cases, indeed, the truncation error
can be made small by keeping a number of states $N_k$ which is much smaller
than the full size of the block Hilbert space [$(2S+1)^{L/2}$ in the present 
case], and can be considered, for all practical purposes, independent of the
system size $L$ at fixed ${\cal R}$. We point out, however, that a small value 
of ${\cal R}$ does not by itself imply a good convergence with respect to the 
truncation of the Hilbert space, first of all because the convergence must be 
directly checked on the observed physical quantity, which may have a peculiar 
dependence on ${\cal R}$, and also because different observables have 
different rates of convergence \cite{caprara}.

The observables are extracted by means of the standard DMRG procedure 
\cite{white}. In this paper we mainly discuss the groundstate energy
$$
E_0(J,H;L,N_k)=\left\langle \left[ {\tilde {\cal H}}\right]^T\right\rangle,
$$
and magnetization
$$
M(J,H;L,N_k)=\left\langle \left[{\tilde S}^z\right]^T\right\rangle,
$$
where the symbol $\langle\cdot\rangle$ denotes the expectation value over the 
DMRG approximate groundstate wavefunction, for the system of size $L$, with 
$N_k$ states retained in the density-matrix truncation procedure, and the 
superscript $T$ means that the corresponding operator has been properly 
truncated onto the superblock basis.

In the following we define the groundstate energy per site in units of $J$,
$e_0=E_0/(JL)$, and the magnetization per site in units of $S$, $m=M/(SL)$,
such that $0\le m\le 1$. We also introduce the notation $h\equiv H/J$, since
$e_0$ and $m$ depend on $J$ and $H$ only through their ratio $h$.

\begin{figure}
\centerline{\psfig{bbllx=80pt,bblly=200pt,bburx=510pt,bbury=575pt,%
figure=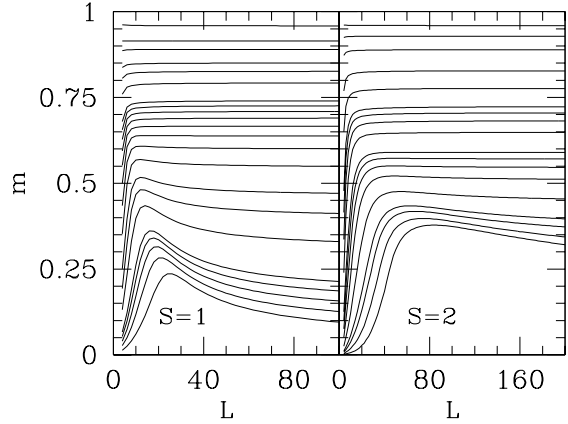,width=70mm,angle=0}}
\vskip 0.5truecm
\caption{Average magnetization $m$ as a function of the length of the 
chain $L$, in the case $S=1$ (left) and $S=2$ (right), for $N_k=50$. 
Notice the different scale on the $L$-axis.
The different curves correspond to increasing values of the magnetic field,
from bottom to top: in the left panel 
$h =$ 0.002, 0.004, 0.006, 0.008, 0.01, 0.02, 0.03, 0.04, 0.06, 0.08, 0.1, 
0.12, 0.14, 0.16, 0.18, 0.2, 0.3, 0.4, 0.5, 0.75, 1.0, 2.0; in the right panel 
$h =$ 0.0001, 0.0003, 0.0006, 0.0009, 0.002, 0.004, 0.006, 0.008, 0.01, 0.02, 
0.03, 0.04, 0.05, 0.1, 0.2, 0.5, 1.0, 2.0. 
\label{bump_fig}
}
\end{figure}

A glance at the magnetization curves as a function of the system size $L$,
for various magnetic fields, and a fixed number of states (e.g., $N_k=50$, see
Fig. \ref{bump_fig}), shows that there are two different behaviors at small
and large field respectively. The convergence of the magnetization to the 
thermodynamic limit is non-monotonic at small $h$, and $m$ is characterized 
by a well-pronounced bump at a characteristic length $L_b$ which increases 
with increasing $S$ and decreases with increasing $h$. This property strongly 
suggests that system sizes smaller than $L_b$ are definitely not 
representative of the thermodynamic limit, and that only for $L \gg L_b$, where
a monotonic convergence to the thermodynamic limit is recovered, the finite 
system is a reasonable representation of the infinite-length chain. For 
$S=1$, at the smallest field we investigated, $h= 0.002$, we found 
$L_b\simeq 25$ (see lowest curve in the left panel of Fig. \ref{bump_fig}). 
This effect is more dramatic as $S$ increases from 1 to 2. Fig. \ref{bump_fig} 
clearly indicates that, for $S=2$, and for the smallest applied field 
$h = 0.0001$, $L_b \simeq 70$, and that, even for $L \simeq 200$, the effect 
of the bump is not negligible, in the sense that the system appears to be far 
from an asymptotic convergence (see lowest curve in the right panel of Fig. 
\ref{bump_fig}). We shall address the physics underlying this behavior in 
Sec. \ref{results}.

It should be clear that the presence of the bump strongly limits the 
possibility to draw reliable conclusions from finite-size DMRG calculations,
unless the latter are pushed to exceedingly large sizes, with huge numerical 
effort. On the other hand, only a careful extrapolation procedure can give 
insights on the actual thermodynamic limit. The infinite-size algorithm is 
certainly suitable for this kind of study.

Within our procedure we extract the thermodynamic limit for any given
number of states $N_k$, by fitting the large-$L$ behavior of the energy
and magnetization as $e_0(h;L,N_k)=e_0^\infty(h;N_k)+A/L$ and 
$m(h;L,N_k)=m^\infty(h;N_k)+B/L$ respectively. The long-living corrections 
are $O(1/L)$ due to the OBC. We explicitly checked that the linear fit in 
$1/L$ at small field is accurate only if $L \gg L_b$. For the smallest 
field (and $S=2$), we had to consider $L \simeq 1600$ to obtain a reliable 
result, and the situation gets worse at larger $S$ \cite{s2}.

Once $e_0^\infty(h;N_k)$ and $m^\infty(h;N_k)$ are obtained, we carefully 
check the convergence with respect to the Hilbert space, for both $m$ and 
$e_0$, by taking the extrapolation to the limit $N_k\to\infty$ in the region 
where the observables display a linear dependence on $1/N_k$ \cite{ising,xy}. 
Up to $N_k=150$ states were needed in the case $S=2$ for the smallest field 
$h=0.0001$, to obtain the linear scaling. We emphasize that this procedure 
explicitly deals with the convergence to the limit of infinite Hilbert space  
of the expectation values of the different observables, and is not equivalent 
to fixing a small truncation error {\it a priori}.

\section{Results and discussion}
\label{results}

In this section we mostly devote our analysis to the case $S=2$, for which
no results in the presence of the magnetic field can be found in the 
literature.

\begin{figure}
\centerline{\psfig{bbllx=80pt,bblly=200pt,bburx=510pt,bbury=575pt,%
figure=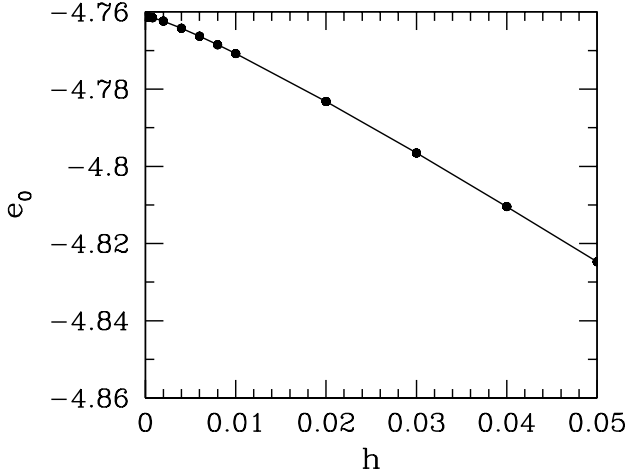,width=70mm,angle=0}}
\vskip 0.75truecm
\caption{Groundstate energy per site in the case $S=2$, extrapolated to the 
thermodynamic limit $L \to \infty$ and to $N_k \to \infty$ (full circles), as 
a function of the magnetic field. The solid line is a guide to the eye.
\label{ene_fig}
}
\end{figure}

\begin{figure}
\centerline{\psfig{bbllx=80pt,bblly=200pt,bburx=510pt,bbury=575pt,%
figure=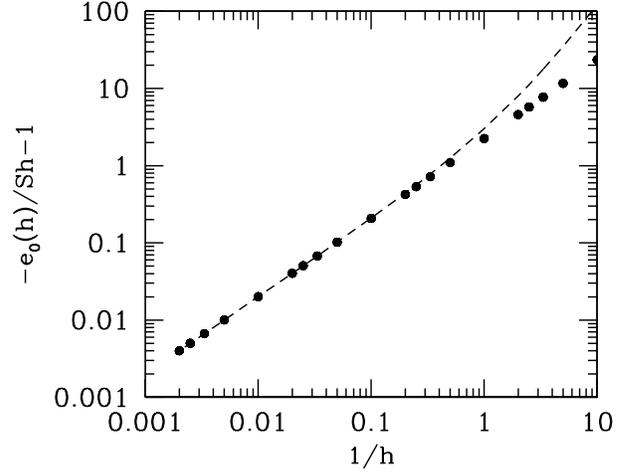,width=70mm,angle=0}}
\vskip 0.75truecm
\caption{Large-field behavior of the groundstate energy per site
in the case $S=2$ (full circles) compared 
with the second-order perturbation theory result, Eq. (\ref{perte}) for
$L\to\infty$ (dashed line). To make the comparison more transparent 
$-e_0(h)/Sh-1$ is plotted against $1/h$, in log-log scale.
\label{ene_fig_zoom}
}
\end{figure}

The groundstate energy per site $e_0$ extrapolated to the infinite-size limit 
($L\to \infty$) and to the full Hilbert space ($N_k\to\infty$), is a smooth 
monotonically decreasing function of the magnetic field, correctly 
interpolating between the zero-field value $e_0(h=0)=-4.76126(1)$, 
and the large-field asymptotic behavior 
$e_0(h\gg 1)\simeq -Sh$. The small-field region is plotted in Fig. 
\ref{ene_fig}, and the large-field region is shown in Fig. \ref{ene_fig_zoom}.
Our result for $e_0(h=0)$ is obtained by an
extrapolation as a function of $h$, and the (relatively) large error
bar is due to this procedure. The value compares reasonably
well with Ref. \cite{s2} ($e_0 = -4.761244(1)$). Our value is slightly
lower, suggesting that our extrapolation procedure is slightly
more efficent that the one in Ref. \cite{s2}.

The magnetization along the magnetic field can be either computed as outlined 
in Sec. \ref{convergency}, by directly evaluating the expectation value of the 
corresponding operator $\tilde S^z$ on the (approximate) DMRG groundstate 
wavefunction, or exploiting the Hellmann-Feynman theorem. The latter implies 
$\langle \tilde S_z \rangle= - {\partial E_0}/ {\partial H}$  when $E_0$ is 
an exact eigenvalue. $m$ may therefore be estimated by a numerical derivative 
of the groundstate energy per site $e_0$ with respect to the field amplitude
$h$. It must be pointed out that when the wavefunction is not an exact 
eigenstate, as in the case of DMRG, the equality does not (necessarily) hold.
However, if the state obtained within DMRG is close to an exact eigenstate,
the Hellmann-Feynman relation is expected to hold within good accuracy.

\begin{figure}
\centerline{\psfig{bbllx=80pt,bblly=200pt,bburx=510pt,bbury=575pt,%
figure=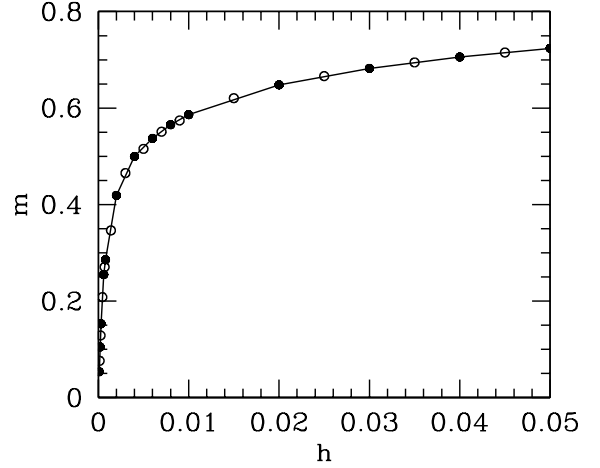,width=70mm,angle=0}}
\vskip 0.75truecm
\caption{Average magnetization per site $m$ in the case $S=2$, extrapolated 
to the thermodynamic limit $L \to \infty$ and to $N _k \to \infty$, as a 
function of the magnetic field. The full circles are direct DMRG measures, 
joined by the solid line as a guide to the eye, while the open dots are 
obtained by numerically differentiating the energy curve (see Fig. 
\ref{ene_fig}) according to the Hellmann-Feynman theorem. 
\label{mag_fig}
}
\end{figure}

The magnetization values obtained following the two procedures are shown in 
Fig. \ref{mag_fig}, and fall on a single regular curve. This remarkable 
agreement represents then a strong evidence of the convergence of our method 
with respect to the truncation of the Hilbert space, that may be used as a 
further check, besides the control of the truncation error $\cal R$, which is 
standard in DMRG\cite{white}, and the extrapolation to $N_k\to\infty$ 
explained in Sec. \ref{convergency}.

\begin{figure}
\centerline{\psfig{bbllx=80pt,bblly=200pt,bburx=510pt,bbury=575pt,%
figure=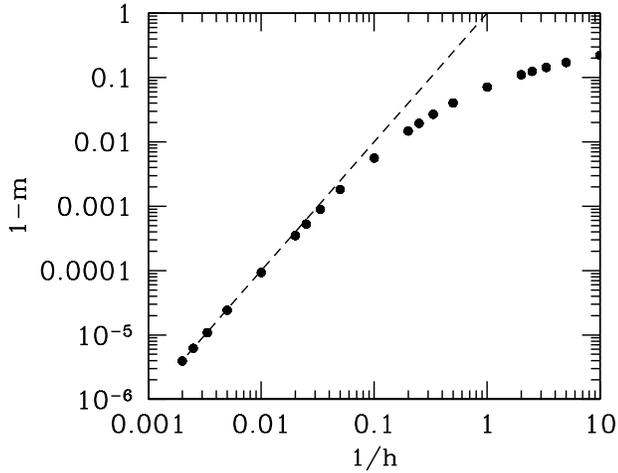,width=70mm,angle=0}}
\vskip 0.75truecm
\caption{Large-field behavior of the average magnetization per site $m$ 
in the case $S=2$ (full circles) compared with the 
perturbation theory result, Eq. (\ref{pertm}) for $L\to\infty$ (dashed line). 
To make the comparison more transparent $1-m$ is plotted against $1/h$, in 
log-log scale.
\label{mag_fig_zoom}
}
\end{figure}

The magnetization is a smooth function of $h$ for all the values of the 
magnetic field, and correctly reproduces both the zero-field limit $m=0$, and 
the $h \gg 1$ limit (shown in Fig. \ref{mag_fig_zoom}), 
where the curve saturates to $1$. Nonetheless, a 
relatively sharp crossover at $h \simeq 0.005$ is signaled in Fig. 
\ref{mag_fig} by a rapid change in the slope of the curve. The 
small-$h$ region is characterized by a spin-liquid behavior, and the 
crossover can be naturally associated with a change in the nature of the 
groundstate from the spin-liquid state to a polarized N\'eel state, induced by 
a gradual freezing of quantum fluctuations. A further support of this 
interpretation comes from the analysis of the curves of $m$ as a function of 
$L$ shown in Fig. \ref{bump_fig}. The small-$h$ region coincides indeed with 
the range in which the bump is observed in the curves. This finding sheds 
light on the physical origin of the bump. The bump can in fact be associated 
with a spin-liquid behavior, since it is observed only in this region.

If we now turn to the magnetization curves of Fig. \ref{bump_fig}, we can 
interpret the non-monotonicity in a simple way. When the system has a 
spin-liquid groundstate, it is characterized by a spin gap, and the 
correlation functions decay with a characteristic lengthscale $\xi$. If 
$L \ll \xi$, the system is too small to display the spin-liquid features, and 
behaves as if $\xi\to\infty$, i.e., similarly to a half-integer spin system, 
which responds more strongly to a small staggered external field. As a result, 
this leads to an ``overshooting'' of the magnetization, that assumes values 
larger than the thermodynamic limit. Only when $L \gg \xi$, the finite-system 
is representative of a spin liquid: the screening of the field becomes 
effective and the magnetization relaxes to its bulk value. This leads to an 
interpretation of $L_b$ as an intrinsic lengthscale related to some 
correlation length $\xi$. A more quantitative study of this relationship 
requires a model for the evolution of the Haldane spin-liquid phase in the 
small-field region, and is currently underway \cite{bigcoll}.

\subsection{The small-field region}
\label{piccoli}

As mentioned above, it is well known that the integer-spin AFM Heisenberg 
chain has a spin-liquid groundstate, characterized by a spin gap, the 
so-called Haldane gap. It is quite natural that the spin-liquid behavior 
extends to small finite fields. As a consequence, at the smallest field 
$h \simeq 0.0001$, the magnetization is linear in the field, as opposed to 
the half-integer spin case, which is critical (i.e. with divergent linear 
susceptibility) at zero field. Our (lower-bound) estimate of the linear 
susceptibility is $\chi(h=0)={\partial m\over\partial h}\vert_{h=0}=1070$, in 
extremely good agreement with the calculations in Ref. \cite{bolognesi}. In 
this region the non-linear $\sigma$-model is therefore the most correct 
field theory. Unfortunately, the Haldane gap decreases by increasing the spin 
$S$, so that the field-driven crossover from the spin liquid to the 
antiferromagnet is expected to occur at an external field which decreases 
with increasing $S$ ($h_c\sim 0.08$ for $S=1$ and $h_c \sim 0.005$, for $S=2$, 
as stated above). For $h > h_c$, the discrepancy with the non-linear 
$\sigma$-model results increases, even if a broad intermediate region exists, 
where the system is no longer a spin liquid, but the groundstate is not yet 
well described as a perturbation of the fully polarized antiferromagnet. In 
this intermediate region, the convergence of the magnetization to the 
thermodynamic limit is monotonic, and the asymptotic value is approached from 
below, contrary to the spin-liquid regime. Finally, a much smoother crossover 
takes place at larger fields, signaling the onset of a large-field 
``perturbative'' region, where $m$ monotonically approaches its limiting 
value from above, as we discuss in Sec. \ref{largefields}. We point out
that the field-driven freezing of the quantum fluctuations is seen in the
gradual simplification of the spectrum of the density matrix, i.e., a 
continuous reduction of the truncation error ${\cal R}$ by increasing $h$ at 
fixed $N_k$.

\subsection{The large-field region}
\label{largefields}

In the limit $h\to\infty$ the groundstate is trivially polarized. In this 
limit the spectrum of the density matrix is therefore dominated by a single 
state, with eigenvalue $w_1=1$, and the truncation error $\cal R$ vanishes
as soon as $N_k>1$. It is therefore reasonable to expect that at large 
magnetic fields, $h\gg 1$, the spectrum of the density matrix will be much 
simpler than in the spin-liquid phase, making the convergence to the full 
Hilbert space relatively easy. As a consequence, the DMRG data are exceedingly 
accurate at $h=1$, even by keeping $N_k \simeq 50$ states (giving 
${\cal R}\sim 10^{-13}$), and at $h=10$, keeping $N_k\simeq 30$ states (giving
${\cal R}\sim 10^{-16}$). Moreover, in this limit we can check our numerical 
data by comparison with the analytical results of standard perturbation 
theory around the $J=0$ fully polarized groundstate. The groundstate 
wavevector for the Hamiltonian (\ref{gauged}) at $J=0$ is 
$|0\rangle = |S\rangle_1 ... |S\rangle_L$, with a groundstate energy
$E_0^{(0)}=-HSL$. As the perturbation Hamiltonian is 
$${\tilde {\cal V}}=-J
\sum_{i=1}^{L-1}[{\tilde S}_i^z {\tilde S}_{i+1}^z+{1\over 2}
({\tilde S}_i^+{\tilde S}_{i+1}^+ + {\tilde S}_i^-{\tilde S}_{i+1}^-)],$$
the groundstate energy up to second order in $J$ is given by
$$E_0\simeq E_0^{(0)}+{\tilde {\cal V}}_{00}
+\sum_{k>0} {|{\tilde {\cal V}}_{k0}|^2 \over E_0^{(0)}-E_k^{(0)}}.$$ 
The excited states which contribute to the sum over $k$ are all degenerate, 
and correspond to wavevectors of the form 
$\vert k\rangle=\vert S\rangle_1 \ldots\vert S-1\rangle_k
\vert S-1\rangle_{k+1}\ldots\vert S\rangle_L$, $k=1,...,L-1$, with 
unperturbed eigenvalues $E_k^{(0)}=-HSL+2H$. As a result,  for $h\gg 1$ we find
\begin{equation}
e_0 \simeq 
-hS\left[1+S{1\over h}+{S\over 2}\left({1\over h}\right)^2\right]
+{S^2\over L}\left(1+{1\over 2h}\right), 
\label{perte}
\end{equation}
where the infinite-size value (first term in the r.h.s.) and the leading
finite-size corrections $O(1/L)$ (second term in the r.h.s.), are explicitly 
shown. The agreement with the DMRG data for $h> 1$ is perfect, as shown in 
Fig. \ref{ene_fig_zoom}.

Analogously one can obtain the magnetization along the field, by calculating 
the groundstate wavefunction up to second order in perturbation theory, and 
then evaluating the expectation value of the operator ${\tilde S}^z$. It is 
however much simpler to make use of the Hellmann-Feynman theorem once more 
and calculate $\langle {\tilde S}^z\rangle =-\partial E_0/\partial H 
\simeq SL-S^2(L-1)/(2h^2)$, obtaining 
\begin{equation}
m\simeq\left( 1-{S\over 2h^2}\right) + {S \over 2h^2 L}.
\label{pertm}
\end{equation} 
We have thus found that the convergence to the thermodynamic limit [the 
first term in the r.h.s. of Eq. (\ref{pertm})] is monotonic and the 
finite-size corrections [the second term in the r.h.s. of Eq. (\ref{pertm})) 
are positive, in perfect agreement with the DMRG data 
(see Fig. \ref{mag_fig_zoom}]. The deviation from the saturation value in the 
thermodynamic limit vanishes as $1/h^2$, in disagreement with the 
$1/\sqrt{h}$ behavior found within the nonlinear $\sigma$-model approach 
\cite{bolognesi}. Since the calculation of Ref. \cite{bolognesi} relies on 
the Haldane {\it ansatz}, the disagreement is expected at large magnetic
fields, where the groundstate is N\'eel-like, rather than spin-liquid-like.

\section{Conclusions}
\label{concl}

We have presented extensive DMRG calculations for integer-spin 
antiferromagnetic Heisenberg chains in an external staggered magnetic field.
The infinite-size DMRG algorithm has proven much more suitable than the 
finite-size version, to extract the thermodynamic properties of these 
systems, due to the presence of sizeable characteristic lengthscales, which
induce anomalous finite-size effects.

We have been able to determine three regions in the phase diagram of the 
$S=2$ chain, as a function of the applied staggered field. The first region, 
for $0 < h \lesssim 0.005$ is the spin-liquid region, in which the spectrum 
is characterized by the Haldane gap, the spin susceptibility is linear in the 
field, with a coefficient $\chi(0) = 1070$, that compares extremely well with 
a non-linear $\sigma$-model calculation \cite{bolognesi}. In this region the 
convergence of the magnetization to the thermodynamic limit is not monotonic 
as the system size $L$ is increased, due to the existence of a characteristic 
lengthscale $\xi$. Thus, quite large systems must be considered to 
correctly describe the spin-liquid phase. Indeed, for $L\ll\xi$ the system
is expected to behave as if $\xi\to\infty$, as it is the case for half-integer
spin, with a strong increase of the magnetization. It is only for $L\gg \xi$
that the screening of the external field in the spin-liquid phase becomes 
effective and the magnetization can relax to its bulk value.

The second region ($0.005 \lesssim h \lesssim 1$) is an intermediate 
crossover region, where the system is no longer a spin liquid, due to the 
field-driven freezing of quantum fluctuations, but the field is not large 
enough to give rise to a linear antiferromagnet. As a result, both the 
non-linear $\sigma$-model (valid in the Haldane phase), and perturbation 
theory around the linear antiferromagnet (obviously valid for $h \to \infty$), 
are not accurate. In this region the magnetization curve monotonically 
approaches the thermodynamic limit, and the limit is reached from below 
(contrary to the Haldane phase).

Finally, for $h > 1$, the magnetic field is large enough to completely freeze 
quantum fluctuations and give rise to an almost saturated N\'eel 
antiferromagnet, that can be well described by perturbation theory in $1/h$. 
As predicted by perturbation theory, the magnetization decreases monotonically
with increasing system size and approaches its thermodynamic limit from above.

A more detailed and quantitative comparison between our numerical results and 
the field-theoretical results of the nonlinear $\sigma$-model is presently in 
progress, aiming to clarify the identification of the characteristic length 
$L_b$ observed in the DMRG data with some correlation length, and the precise 
region of validity of the analytical approaches \cite{bigcoll}.

\begin{acknowledgments}

It is a pleasure to thank F. Becca for a substantial contribution to the 
improvement of the DMRG code. We also thank L. Capriotti, E. Ercolessi, G. 
Morandi, F. Ortolani, P. Pieri, M. Roncaglia, for stimulating discussions.
 
\end{acknowledgments}

\end{document}